\documentclass[onecolumn,conference]{IEEEtran}

\usepackage{bbm}
\usepackage{bm}
\usepackage{cite}
\usepackage{amsmath,amssymb,amsfonts}
\usepackage{algorithm}
\usepackage{algpseudocode}
\usepackage{graphicx}
\newtheorem{remark}{Remark}

\def\N{\mathcal{N}}

\def\Pc{\mathcal{P}}

\def\rhot{\tilde{\rho}}

\def\Rt{\widetilde{R}}

\def\0{\bm{0}}
\def\1{\bm{1}}

\newcommand{\he}[1]{\parbox{0pt}{\rule{0pt}{#1}}}
\newtheorem{propo}{Proposition}

\begin{document}

\title{Digital Fairness Algorithms \\
for Satellite Uplink NOMA}

\author{\IEEEauthorblockN{Giorgio Taricco}
\IEEEauthorblockA{\textit{Politecnico di Torino -- DET, Torino Italy} \\
E-mail: giorgio.taricco@polito.it}
}
\maketitle

\begin{abstract}
Achieving digital fairness by using NOMA is one of the more pressing issues in modern wireless communication systems for 5G/6G networks.
This is particularly true in the case of satellite uplink systems supporting a population of IoT wireless devices scattered in a wide coverage area.
In this scenario, the variability of the link budget across space and time increases the challenges of preventing a situation where only a subset of network users can transmit while others are left unable to do so.
This work investigates the characteristics of an uplink NOMA system with the goal of equalizing the achievable rate of the IoT network subscribers.
Within the context of single-slot NOMA, two key outcomes are achieved: the determination of the optimal SIC ordering at the receiver and the exploration of power moderation, coordinated by the receiver, to maximize the minimum user rate.
In the context of multi-slot NOMA, which is particularly relevant to the satellite scenario under consideration, a user rate equalization algorithm is proposed and its performance is analyzed numerically.
The trade-off between network performance, measured in terms of user rates, and complexity, determined by the number of SIC steps implemented at the receiver, is thoroughly evaluated for the satellite scenario under consideration.
\end{abstract}

\begin{IEEEkeywords}
5G,
6G,
Uplink NOMA,
Successive Interference Cancellation,
LEO satellite networks,
Digital fairness.
\end{IEEEkeywords}

\section{Introduction}

Non-Orthogonal Multiple Access (NOMA) has emerged as a key enabling technology for next-generation wireless communication systems, including 5G and beyond. 
Unlike traditional orthogonal multiple access (OMA) schemes, NOMA allows multiple users to share the same time-frequency resources, thereby improving spectral efficiency and connectivity. 
In the uplink scenario, NOMA introduces unique challenges and opportunities, particularly in achieving fairness among users with varying channel conditions and power levels. 
MAX-MIN fairness, which aims to maximize the minimum performance metric (usually the achievable information rate) among all users, has been widely studied as a fairness criterion in this context.
For example,
Zhang \textit{et al.} \cite{zhang2023maxmin} explored MAX-MIN fairness for uplink NOMA systems under finite blocklength constraints.
%Their work focuses on the joint optimization of power control and transmission rates to enhance the minimum effective throughput while ensuring quality-of-service (QoS) requirements. %They propose an iterative algorithm based on successive convex approximation to solve the non-convex optimization problem efficiently. The results demonstrate significant improvements in fairness and overall system performance.
Xu and Clerckx \cite{xu2024maxmin} proposed an uplink MIMO RSMA framework for short-packet communications with perfect channel state information.
%It optimized precoders and combiners under Max-Min Fairness and Finite Blocklength constraints using Alternating Optimization and Successive Convex Approximation.
%A low-complexity decoding order was also designed. Simulations showed RSMA outperforms traditional schemes like SDMA and NOMA, offering better Max-Min Fairness performance, robustness to network loads, and improved throughput.
Liu \textit{et al.} \cite{liu2019fairness} address the near-far problem in wireless-powered communication networks using a fairness-aware NOMA-based scheduling scheme to enhance max-min fairness. 
%Using order statistic theory, they find a scheme to form optimal user groups for NOMA transmission.

Focusing on uplink NOMA satellite communication systems, it has been observed that achieving fairness among users with diverse channel conditions, transmission powers, and geometrical positions poses unique challenges.
The inherent differences in user distances from the satellite, combined with non-uniform beamforming gains and frequency reuse patterns, exacerbate the disparity in signal quality, making resource allocation and interference management critical to ensuring equitable access to communication services.
Additionally, the high mobility of satellites in low Earth orbit (LEO) introduces dynamic changes in user-satellite geometry, necessitating adaptive algorithms to maintain fairness and quality-of-service guarantees in real-time.
%The integration of Non-Orthogonal Multiple Access (NOMA) in uplink satellite and related IoT communication scenarios has been extensively studied to address challenges like fairness, spectral efficiency, and dynamic user associations.
In this area,
Wang \textit{et al.} \cite{wang2020noma} proposed a NOMA-enabled framework for multi-beam satellite systems to overcome the mismatch between offered capacity and requested traffic. 
%Their joint optimization algorithm integrates power allocation, decoding order determination, and terminal-timeslot assignment, improving the Offered Capacity to Requested Traffic Ratio (OCTR). 
This study highlights the applicability of NOMA in satellite networks to balance load and enhance fairness among geographically dispersed users.
From a different standpoint,
Ahsan \textit{et al.} \cite{ahsan2020resource} explored the use of reinforcement learning in uplink NOMA for Internet of Things (IoT) networks. 
%Their approach leverages Deep Reinforcement Learning (DRL) and SARSA algorithms to optimize resource allocation dynamically, achieving superior system throughput and adaptability. 
This is relevant in satellite IoT use cases where diverse traffic demands require flexible and intelligent solutions.
In a related context,
Nauman \textit{et al.} \cite{nauman2024efficient} examined a vehicular-aided heterogeneous network (HetNet) with High Altitude Platforms (HAPs) and uplink NOMA. 
%They introduced an advanced user association and resource allocation framework tailored for vehicular mobility, enhancing both spectral efficiency and fairness. 
Their results underscore the potential of combining NOMA with advanced network topologies like HAPs and vehicular HetNets to address mobility-induced challenges.
%In the framework of uplink NOMA, achieving MAX-MIN fairness and determining the optimal Successive Interference Cancellation (SIC) order is crucial to ensure efficient utilization of resources while providing fairness to users with varying channel conditions.
In the framework of MAX-MIN fairness for uplink NOMA, Gao \textit{et al.} \cite{gao2017} provide evidence that the optimum SIC order corresponds to decoding the signals from the stronger to the weaker to minimize a certain outage probability (which is not equivalent to the case considered here).

\subsection{Organization}
The study is organized as follows.
Section \ref{sec:scenario} presents a realistic satellite beamspot scenario and investigates the properties of the SNR at the receiver corresponding to randomly distributed IoT user terminals with uplink NOMA.
Section \ref{sec:uplinknoma} provides a definition of a time-slotted uplink NOMA system and reviews the expressions for the user rates achievable by Successive Interference Cancellation along with the overall sum-rate.
Section \ref{sec:SIC.order} addresses the optimum SIC order for a specific set of SNRs.
Section \ref{sec:reducepower} introduces a novel approach to the optimization of the maximum minimum rate based on the moderation of the transmitted user powers.
Section \ref{sec:multislot} addresses the typical multi-slot scenario experienced by the majority of wireless communication systems, where the channel remains approximately constant during time intervals called time-slots and the performance over a set of successive time-slots is targeted.
The section proposes an algorithm to equalize the user rates over the multi-slot horizon.
Section \ref{sec:numerical} provides a numerical analysis of the scenario introduced in Section \ref{sec:scenario} with the goal of characterizing the performance versus complexity trade-off of the network.
Finally, Section \ref{sec:conclusions} summarizes the contribution of the work and provides concluding remarks.

\section{Satellite beamspot scenario} \label{sec:scenario}

The scenario considered is inspired to the Starlink satellite network, developed by SpaceX, with some simplifying assumptions.
This satellite network is based on a LEO satellite mega-constellation designed for global, low-latency internet coverage.
The system achieves its performance goals through the structured organization of orbital shells, ground station integration, and inter-satellite communication.
Starlink satellites are organized into multiple orbital shells at varying altitudes and inclinations. 
These shells provide global coverage, with the satellites positioned to ensure continuous visibility over different regions of the Earth. 
The key parameters including altitude, inclination, and the number of satellites in each shell are illustrated in Table \ref{tab:orbital_structure} \cite{rinaldi2020,azari2022,capez2024}.

\begin{table}
\def\thetable{\arabic{table}}
\centering
\caption{Orbital structure of Starlink satellites.}
\begin{tabular}{|c|c|c|c|c|c|}
\hline
\textbf{Shell} & \textbf{Altitude (km)} & \textbf{Incl. (°)} & \textbf{Planes} & \textbf{Sats/Plane} & \textbf{Total} \\ \hline
1 & 540 & 53.3 & 72 & 22 & 1584 \\ \hline
2 & 550 & 53.0 & 72 & 22 & 1584 \\ \hline
3 & 560 & 97.6 &  6 & 58 &  348 \\ \hline
4 & 560 & 97.6 &  4 & 43 &  172 \\ \hline
5 & 570 & 70.0 & 36 & 20 &  720 \\ \hline
\end{tabular}
\label{tab:orbital_structure}
\end{table}

The satellites in each orbital shell move in nearly circular paths, completing one orbit around Earth in $90$ minutes.
This corresponds to an angular speed of $$\frac{360^\circ}{90\times60\text{ s}}=0.067^\circ/\text{ s}.
$$
The typical 3-dB beam-width of a Starlink satellite antenna is $3.5^\circ$, corresponding approximately to a central Earth angle of $0.3^\circ$ and a satellite passage time of $4.5$ s.
The satellite gain $G(\psi)$ (in dBi) can be modeled according to the following analytical approximation \cite{ITU-R_S.1528}:
\begin{align}\label{eq:ITU.Gain}
G(\psi)=
\begin{cases}
G_\mathsf{max}-3(\psi/\psi_\mathsf{b})^\alpha&\psi\le a\psi_\mathsf{b}\\
G_\mathsf{max}+L_\mathsf{L}-20\log_{10}z&a\psi_\mathsf{b}<\psi\le (b/2)\psi_\mathsf{b}\\
G_\mathsf{max}+L_\mathsf{L}&(b/2)\psi_\mathsf{b}<\psi\le  b\psi_\mathsf{b}\\
X-25\log_{10}\psi&b\psi_\mathsf{b}<\psi\le Y\\
L_\mathsf{F}&Y\le\psi<90^\circ\\
L_\mathsf{B}&90^\circ\le\psi\le180^\circ
\end{cases}
\end{align}
Here,
$\psi$ is the off-axis angle from the satellite (in degrees),
$\psi_\mathsf{b}$ is one half of the 3-dB angle (in degrees),
$G_\mathsf{max}$ is the maximum gain in the main lobe (dBi),
$X=G_\mathsf{max}+L_\mathsf{L}+25\log_{10}(b\psi_\mathsf{b})$,
$Y=b\psi_\mathsf{b}^{0.04(G_\mathsf{max}+L_\mathsf{L}-L_\mathsf{F})}$,
$L_\mathsf{L}$ is the near-in side-lobe level,
$L_\mathsf{F}$ is the 0-dBi far-out side-lobe level,
$L_\mathsf{B}$ is the back-lobe level, and
$z$ is the major/minor axis ratio.

\begin{figure}
\centering
\includegraphics[width=.8\linewidth]{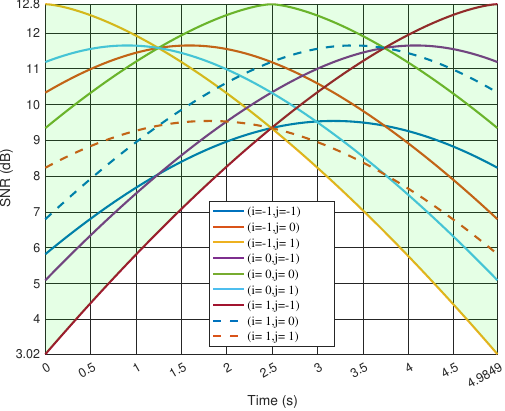}
\caption{SNR diagrams corresponding to user locations
$(\lambda_0+i\delta_\mathsf{lat},\varphi_0+j\delta_\mathsf{lon})$ for $i,j\in\{-1,0,1\}$ in latitude/longitude coordinates.
The light green area corresponds to the SNR range over the coverage area.}
\label{fig:SNR}
\end{figure}

Consider a scenario where the satellite orbits at a distance of $550$ km from the Earth surface and communicates in the Ku-band at $14$ GHz.
The reference antenna is pointed to the satellite nadir and its maximum gain is $G_\mathsf{max}=36$ dBi.
The IoT terminals transmit at a power of $P_\mathsf{tx}=10$ W, with an antenna gain of $3$ dBi.
Free-space path loss is used for signal attenuation, based on the distance between the satellite and the IoT terminals.
The noise power is calculated on a bandwidth of $10$ MHz and a noise temperature of $290$ K, yielding a noise power spectral density of $-174$ dBm/Hz.
A line-of-sight (LOS) model is used, assuming no obstructions or atmospheric effects (\textit{clear-sky} conditions).
The received signal power from each IoT terminal is calculated by the standard Friis equation \cite{balanis2016antenna}:
\begin{equation}
P_{\mathsf{rx}} = G_{\mathsf{term}}G_{\mathsf{sat}}(\psi)
\bigg(\frac{\lambda}{4\pi d}\bigg)^2P_{\mathsf{tx}}
\end{equation}
$P_{\mathsf{tx}}$ is the transmitted power from the IoT terminal,
$G_{\mathsf{term}}$ is the IoT terminal's antenna gain (0 dBi for an isotropic antenna),
$G_{\mathsf{sat}}(\psi)$ is the satellite antenna gain, modeled according to eq.\ \eqref{eq:ITU.Gain},
$\lambda$ is the carrier wavelength, and
$d$ is the propagation distance.
The received Signal-to-Noise Ratio (SNR) is:
\begin{equation}
\mathsf{SNR} = \frac{P_{\mathsf{rx}}}{P_{\text{noise}}} = \frac{P_{\mathsf{rx}}}{kT_0B}.
\end{equation}
Here, $k=1.38\cdot10^{-23}$ J/K is the Boltzmann constant, $T_0=290$ K is the Earth average temperature and $B=10$ MHz is the bandwidth.
The coverage region is defined by the latitude/longitude ranges:
$$
\lambda\in(\lambda_0-\delta_\mathsf{lat},\lambda_0+\delta_\mathsf{lat}),\quad
\varphi\in(\varphi_0-\delta_\mathsf{lon},\varphi_0+\delta_\mathsf{lon})
$$
with
$\delta_\mathsf{lat}=0.1^\circ$ and
$\delta_\mathsf{lon}=0.1\cdot\tan(53^\circ)=0.133^\circ$.
The satellite nadir passes through the center of the region $(\lambda_0,\varphi_0)$ with an inclination of $53^\circ$ with respect to the equatorial plane, and the passage time is $\frac{0.2^\circ}{\cos(53^\circ)\cdot0.067^\circ/ \mathrm{s}}=4.98$ s.
The received SNR corresponding to the nine user locations characterized by their latitude/longitude coordinates $(\lambda_0+i\delta_\mathsf{lat},\varphi_0+j\delta_\mathsf{lon})$ for $i,j\in\{-1,0,1\}$ is illustrated in Fig.\ \ref{fig:SNR}.
The scenario is further detailed by including a total of $N$ users within the coverage area and dividing the time frame into $T$ time slots, during which the received SNRs remain nearly constant (\textit{e.g.}, setting $T=100$ limits the SNR variability within each slot to approximately $0.1$ dB or less).

\section{Uplink NOMA}\label{sec:uplinknoma}

Consider a multiuser communication system where $N$ IoT user terminals transmit to a LEO satellite by using a Non Orthogonal Multiple Access (NOMA) technique.
The terminals, labeled by $n=1,\ldots,N$, transmit over a sequence of $T$ time slots labeled by $t=1,\ldots,T$.
The channel gain is approximately constant over the time slots so that the received signal at the satellite can be characterized by the following equation:
\begin{align}
    Y[t]=\sum_{n=1}^NH_n[t]X_n[t]+Z_n[t],\quad t=1,\ldots,T.
\end{align}
Define the satellite receiver SNRs:
\begin{align}
    \rho_n[t]\triangleq\frac{|H_n[t]|^2P_n[t]}{P_z}
\end{align}
Here, $P_n[t]$ is the transmitted power from the $n$-th IoT user terminal during the $t$-th time slot, and $P_z$ is the receiver noise power.
Following the standard approach of superposition encoding and Successive Interference Cancellation (SIC), which is well developed and described in the literature (see, \textit{e.g.}, \cite{tse2005fundamentals}), it can be shown that the achievable rates of the multiple-access multiuser communication system are given by:
\begin{align}\label{eq:Rates}
    R_n[t] = \log_2\bigg(1+\frac{\rho_n[t]}{1+\sum_{m>n}\rho_m[t]}\bigg)
\end{align}
for $n=1,\ldots,N,t=1,\ldots,T$.
For simplicity, we disregard possible frequency selectivity and assume that one time slot is sufficiently long to allow the transmission of one codeword.
The rate $R_n[t]$ is achieved by decoding the codeword corresponding to user $n$ during the time slot $t$ after decoding the codewords corresponding to the users $1,\ldots,n-1$ and, after each decoding, re-encoding the information symbols and removing their contribution from the received signal.
It is plain to see that the sum-rate
\begin{align}
R_\text{sum}[t]=\sum_{n=1}^NR_n[t]=\log_2(1+\rho_1[t]+\ldots+\rho_N[t])
\end{align}
is achieved by every possible SIC order.
The achievable rates corresponds to a corner point of the polytope representing the $N$-dimensional capacity region \cite{tse2005fundamentals}.
However, in order to limit the spreading of the achievable rates, it is better to order the SIC from the strongest to the weakest user (i.e., from the user with highest to lowest SNR), as discussed in detail in the following section \ref{sec:SIC.order}.

The implementation of the NOMA architecture requires full knowledge of the Channel State Information at the Receiver (CSIR), corresponding to the channel gains $H_n[t]$.
Moreover, the receiver (i.e., the satellite) must be able to coordinate the user transmission powers $P_n[t]$ in order to optimize the target rate according to a fairness criterion.

\subsection{Optimum SIC order (single slot)}\label{sec:SIC.order}

Adopting the MAX-MIN fairness criterion, the optimum SIC order consists of decoding the signals following the decreasing SNR order.
This preferred order is mentioned without proof in \cite[Sec.~6]{tse2005fundamentals} as the \textit{natural} order for the uplink multiuser channel.
This is stated precisely in the following proposition.

\begin{propo}\label{th:SIC}
For a given set of SNRs, $\{\rho_1,\ldots,\rho_N\}$, the maximum minimum achievable rate is obtained when the SNR sequence is nonincreasing.
\end{propo}

\begin{IEEEproof}
See App.\ \ref{app:SIC}.
\end{IEEEproof}

\subsection{Optimization by power moderation (single slot)}
\label{sec:reducepower}

If the SIC order is chosen according to the MAX-MIN fairness criterion, one can maximize the minimum rate by reducing the transmitted powers as is stated in the following proposition.
\begin{propo}\label{th:reducepower}
For a given set of nonincreasingly ordered SNR upper bounds, $\{\rho_1,\ldots,\rho_N\}$, the maximum minimum achievable rate is obtained when the SNR sequence satisfies the following conditions.
\begin{itemize}
\item For $n=1,\ldots,N$, solve the equations
$$\rho_n=\varphi_n(R)\triangleq2^{(N-n+1)R}-2^{(N-n)R},$$
and let $\Rt_n$ be the unique solutions.
\item Let $\Rt\triangleq\min\{\Rt_1,\ldots,\Rt_n\}.$
\item The corresponding SNRs are given by
$$\rhot_n=\varphi_n(\Rt).$$
\end{itemize}
\end{propo}

\begin{IEEEproof}
See App.\ \ref{app:reducepower}.
\end{IEEEproof}
\begin{remark}
It is important to note that, as stated in Proposition \ref{th:reducepower}, certain user transmission powers can be reduced (by a factor $\rho_n/\rhot_n$) while simultaneously increasing the minimum rate.  
This paves the way to a coordinated approach for enhancing MAX-MIN fairness in a NOMA uplink, only by moderating the users' transmission powers.
\end{remark}

\subsection{Multi-slot optimization}\label{sec:multislot}

While Sections \ref{sec:SIC.order} and \ref{sec:reducepower} focused on the MAX-MIN fairness optimization of an uplink NOMA transmission system during a single instance of received SNRs, the time variability of the channel conditions suggests that a different type of objective function must be taken into account in order to consider the performance over a longer time span.

Recalling the scenario introduced in Section \ref{sec:scenario}, $N$ IoT user terminals accessing a NOMA uplink transmit with a maximum power $P_\mathsf{tx}$, which can be decreased according to the satellite instructions.
We define a \textit{feasible power region}
\begin{align}
\Pc\triangleq\{P_n[t]\le P_\mathsf{tx},n=1,\ldots,N,t=1,\ldots,T\}.
\end{align}
Since the IoT user terminals have the goal to exploit the achievable information rate to the satellite during its passage on the coverage area, the following optimization problem based on a MAX-MIN fairness criterion can be formulated:
\begin{align}
\mathop{\mathrm{maximize}}_{P_n[t]\in\mathcal{P}}
\min_{1\le n\le N}\sum_{t=1}^TR_n[t].
\end{align}
Here, $R_n[t]$ is defined according to \eqref{eq:Rates}.
During each time-slot, the SIC order at the receiver and the transmission powers are chosen to achieve the MAX-MIN fairness criterion as illustrated in Sections \ref{sec:SIC.order} and \ref{sec:reducepower}.

Complexity limitations related to the implementation of the SIC suggest that only a fixed number, $N_\mathsf{SIC}$, of users be allowed to transmit during each time slot
\footnote{
The other users must not be transmitting during the time slot.
This requires network coordination incompatible with grant-free transmission \cite{shahab2020}.
}.
Algorithm \ref{algo:1} illustrates a multi-slot uplink NOMA rate optimization scheme based on the results from Proposition \ref{th:SIC} or \ref{th:reducepower}.

\begin{algorithm}\label{algo:1}
\caption{Uplink NOMA multi-slot optimization}
\begin{algorithmic}[1]
\State \textbf{Input:}
$\rho_n[t],n=1,\ldots,N,t=1,\ldots,T$ and $N_\mathsf{SIC}$
\State \textbf{Output:} $\rhot_n[t],n=1,\ldots,N,t=1,\ldots,T$
\State \textbf{Initialize} the cumulative rates $R_n=0,n=1,\ldots,N$
\For{$t=1,\ldots,T$}
\State Select the $N_\mathsf{SIC}$ users with minimum cumulative rates
\State Let $\N_\mathsf{SIC}[t]$ denote this set of users
\State Apply Prop.\ \ref{th:SIC} or \ref{th:reducepower} to calculate the rates $\Rt_n[t]$
\State for all $n\in\N_\mathsf{SIC}[t]$
\State Update the cumulative rates $R_n$ by adding the $\Rt_n[t]$
\State for all $n\in\N_\mathsf{SIC}[t]$
\EndFor
\State \textbf{return} $\N_\mathsf{SIC}[t],\rhot_n[t],R_n$ for $n=1,\ldots,N,t=1,\ldots,T$
\end{algorithmic}
\end{algorithm}

\section{Numerical Results}\label{sec:numerical}

This section presents numerical results concerning the information rates achieved by the satellite IoT user terminals populating the scenario defined in Section \ref{sec:scenario} under the following operating assumptions.
\begin{itemize}
\item $N=256$ IoT user terminals are considered, distributed over a $16\times16$ square grid over the coverage area.
\item The satellite cycle time of about $5$ seconds is partitioned into $T=100$ time slots of $50$ ms (sufficiently long to accommodate one codeword at the bandwidth considered).
\item Algorithm \ref{algo:1} is employed to calculate the achievable rates for the available bandwidth $B=10$~MHz, using a periodic extension of the satellite cycle time over a sufficiently large number of repetitions ($N_\mathsf{rep}=100$).
\item The impact of a random permutation of the time slots within one satellite cycle is also considered in the results.
\item The results compare the rates achieved by using or not the \textit{power moderation} scheme described in Section~\ref{sec:reducepower}.
\end{itemize}

\begin{figure}\centering
\includegraphics[width=0.8\linewidth]{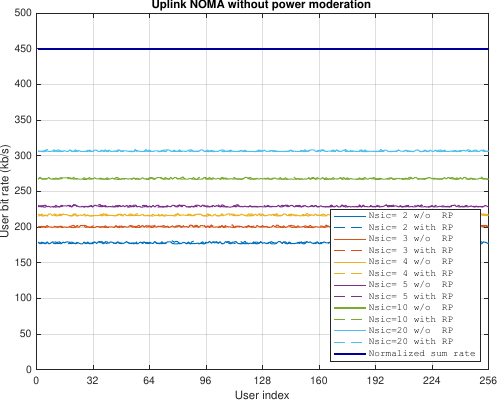}
\caption{User achievable rates obtained by applying Algorithm \ref{algo:1} without power moderation in the scenario considered.
The upper limit labeled "Normalized sum-rate''
corresponds to $\frac{1}{T}\sum_t\log_2(1+\sum_n\rho_n[t])\cdot B$.
}
\label{fig:rates_no_pow_mod}
\end{figure}

\begin{figure}\centering
\includegraphics[width=0.8\linewidth]{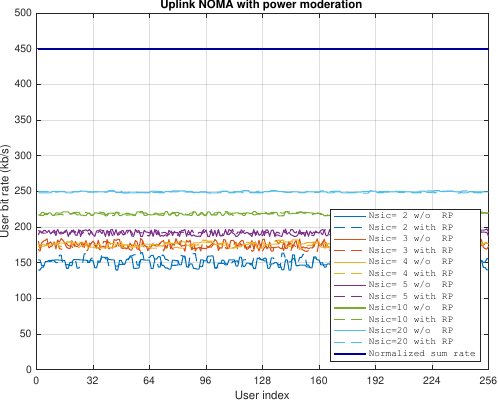}
\caption{Same as Fig.\ \ref{fig:rates_no_pow_mod} but with power moderation.}
\label{fig:rates_pow_mod}
\end{figure}

\begin{figure}\centering
\includegraphics[width=0.8\linewidth]{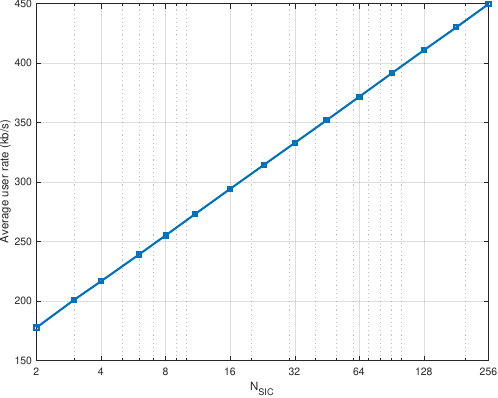}
\caption{Plot of the user achievable rate versus the SIC order $N_\mathsf{SIC}$ for the scenario considered.}
\label{fig:rates_vs_sic}
\end{figure}

Figs. \ref{fig:rates_no_pow_mod} and \ref{fig:rates_pow_mod} illustrate the achievable rates without and with power moderation, respectively.
In both cases, the results corresponding to $N_\mathsf{SIC}=2,3,4,5,10,$ and $20$ are reported.
For each value of $N_\mathsf{SIC}$, the impact of a random permutation of the processing order in Algorithm \ref{algo:1} is assessed.
Finally, both figures reports the bit rate corresponding to the average sum-rate over the $T$ time slots.
This value corresponds to the implementation of the complete SIC scheme ($N_\mathsf{SIC}=N$).
Several comments on these results are in order.
\begin{itemize}
\item The impact of permutation of the time slots is negligible in all cases, so that it is convenient processing the time slots in their natural temporal order.
\item The variations of the user achievable rates are very limited and decrease as the SIC order $N_\mathsf{SIC}$ increases.
Therefore, \textit{fairness} is achieved by Algorithm \ref{algo:1}.
\item The dependence of the user achievable rates on the SIC order is illustrated in Fig.\ \ref{fig:rates_vs_sic}.
The curve shows a linear increase of the rate with the logarithm of $N_\mathsf{SIC}$.
This enables a trade-off between network performance (user rate) and receiver complexity (determined by the SIC order $N_\mathsf{SIC}$).
\item Although the MAX-MIN fairness is enhanced by applying the power moderation approach described in Section \ref{sec:reducepower}, this benefit diminishes over time since Algorithm \ref{algo:1} achieves rate equalization without requiring additional measures.
Moreover, power moderation reduces the sum-rate and therefore the average user rates.
\end{itemize}

\section{Conclusions}\label{sec:conclusions}

Fairness in Non-Orthogonal Multiple Access (NOMA) systems is a critical challenge, particularly in 5G/6G satellite uplinks where IoT devices face variable link budgets. 
This variability often limits some users from transmitting effectively, creating inequities in network performance.
This work analyzed uplink NOMA with the goal of rate equalization among IoT users. For single-slot NOMA, two key results were achieved:
$1.$ Optimal Successive Interference Cancellation (SIC) ordering at the receiver.
$2.$ Power moderation to maximize the minimum user rate and improve fairness.
In multi-slot NOMA, relevant to satellite scenarios, a novel algorithm for rate equalization was proposed and analyzed. The trade-off between user rates and system complexity, particularly SIC steps, was assessed.
These findings demonstrate the potential for fairer and more efficient uplink NOMA systems, advancing the design of next-generation wireless networks.
Future work could address real-time conditions such as the transient behavior and the latency.

\appendices
\section{Proof of Proposition \ref{th:SIC}}\label{app:SIC}

Consider an initial SNR sequence such that $\rho_a<\rho_b$ for some indexes $a<b$.
Then, exchanging these two SNR values has the following effects on the rates:
\begin{itemize}
\item For $i<a$ or $i>b$, the rates $R_i$ remain unchanged. This is because the expressions $\rho_i/(1+\rho_{i+1}+\ldots+\rho_N)$ are unaffected by the swapping of $\rho_a$ and $\rho_b$.
\item For $i=a$, the rate $R_a$ changes as:
\begin{align*}
&R_a=\log_2\bigg(1+\frac{\rho_a}{1+\rho_{a+1}+\ldots+\rho_N}\bigg)
\mapsto\\
&R_a'=\log_2\bigg(1+\frac{\rho_b}{1+(\rho_a-\rho_b)+\rho_{a+1}+\ldots+\rho_N}\bigg)
\end{align*}
\item For $a<i<b$, the rate $R_i$ as:
\begin{align*}
&R_i=\log_2\bigg(1+\frac{\rho_i}{1+\rho_{i+1}+\ldots+\rho_N}\bigg)
\mapsto\\
&R_i'=\log_2\bigg(1+\frac{\rho_i}{1+(\rho_a-\rho_b)+\rho_{i+1}+\ldots+\rho_N}\bigg)
\end{align*}
Since $\rho_a<\rho_b$, all these rates increase.
\item For $i=b$, the rate $R_b$ changes as:
\begin{align*}
&R_b=\log_2\bigg(1+\frac{\rho_b}{1+\rho_{b+1}+\ldots+\rho_N}\bigg)
\mapsto\\
&R_b'=\log_2\bigg(1+\frac{\rho_a}{1+\rho_{b+1}+\ldots+\rho_N}\bigg)
\end{align*}
\end{itemize}
Summarizing the previous results, $$\min\{R_a,R_b\}<\min\{R_a',R_b'\}$$ since
\begin{align*}
\frac{\rho_a}{1+\rho_{a+1}+\ldots+\rho_N} &<
\frac{\rho_a}{1+\rho_{b+1}+\ldots+\rho_N} \\
\frac{\rho_b}{1+\rho_{b+1}+\ldots+\rho_N} &<
\frac{\rho_b}{1+(\rho_a-\rho_b)+\rho_{a+1}+\ldots+\rho_N}
\end{align*}
Merging with the inequalities for $a<i<b$, we get
\begin{align*}
\min_{a\le i\le b}R_i < \min_{a\le i\le b}R_i'.
\end{align*}
Finally, considering the fact that the rates remain unchanged for $i<a$ and $i>b$, the minimum rate increases if it was attained when $a\le i\le b$ and doesn't change otherwise.
Therefore, whenever $\rho_a<\rho_b$, the two SNRs can be exchanged and the minimum rate does not decrease.
This implies that the maximum minimum rate is achieved when the SNR sequence is nonincreasing.
Table \ref{tab:SIC.th} provides an illustrative example of the proof concept.
\begin{table}
\centering
\caption{Illustrative example \\for the proof of Proposition \ref{th:SIC}.}
\begin{tabular}{|cccc|c|}\hline
\multicolumn{4}{|c|}{SNRs}& $R_\mathsf{min}$
\he{4mm}
\\\hline
\he{3.5mm}1 & 3 & 6 & 10 & 0.0704\\\hline
\he{3.5mm}10 & 3 & 6 & 1 & 0.4594\\\hline
\he{3.5mm}10 & 6 & 3 & 1 & 0.9329\\\hline
\end{tabular}
\label{tab:SIC.th}
\end{table}

\section{Proof of Proposition \ref{th:reducepower}}
\label{app:reducepower}

The definition of $\Rt$ in the statement of Proposition \ref{th:reducepower} implies that $\Rt\le\Rt_n$ for all $n=1,\ldots,N$.
Then, $\rhot_n\le\rho_n$ for all $n=1,\ldots,N$.
Moreover, since $\rhot_n=2^{(N-n+1)\Rt}-2^{(N-n)\Rt}$,
$$\log_2\bigg(1+\frac{\rhot_n}{1+\sum_{m>n}\rhot_m}\bigg)=\Rt.$$
Let $R_\mathsf{min}$ be the minimum rate corresponding to the original SNR sequence.
Since the goal is to show that $\Rt\ge R_\mathsf{min}$, assume, on the contrary, that $\Rt<R_\mathsf{min}$, or equivalently that $\Rt<R_n$ for all $n=1,\ldots,N$.
For $n=N$, the inequality implies that $\rhot_N<\rho_N$, which is satisfied.
For $n=N-1$, the inequality implies that 
$$
\frac{\rho_{N-1}}{1+\rho_N}
>
\frac{\rhot_{N-1}}{1+\rhot_N}
>
\frac{\rhot_{N-1}}{1+\rho_N}
\implies
\rhot_{N-1}<\rho_{N-1}.
$$
For $n=N-2$,
$$
\frac{\rho_{N-2}}{1+\rho_{N-1}+\rho_N}
>
\frac{\rhot_{N-2}}{1+\rhot_{N-1}+\rhot_N}
>
\frac{\rhot_{N-2}}{1+\rho_{N-1}+\rho_N},
$$
which implies that
$$
\rhot_{N-2}<\rho_{N-2}.
$$
Proceeding in this way down to the last case of $n=1$, we can see that the inequality $\Rt<R_\mathsf{min}$ implies that $\rhot_n<\rho_n$ for all $n=1,\ldots,N$.
However, since $\Rt=\min\{\Rt_1,\ldots,\Rt_N\}$, $\Rt=\Rt_\nu$ for at least one $\nu\in\{1,\ldots,N\}$.
Then,
$$
\rho_\nu=2^{(N-\nu+1)\Rt_\nu}-2^{(N-\nu)\Rt_\nu}=2^{(N-\nu+1)\Rt}-2^{(N-\nu)\Rt}=\rhot_\nu,
$$
contradicting the consequences of the assumption $\Rt<R_\mathsf{min}$ and proving that $\Rt\ge R_\mathsf{min}$, which is the statement of Proposition \ref{th:reducepower}.

\newpage
\bibliographystyle{IEEEtran}
\bibliography{ISIT2025_NOMA.bib}

% Generated by IEEEtran.bst, version: 1.14 (2015/08/26)
\begin{thebibliography}{10}
\providecommand{\url}[1]{#1}
\csname url@samestyle\endcsname
\providecommand{\newblock}{\relax}
\providecommand{\bibinfo}[2]{#2}
\providecommand{\BIBentrySTDinterwordspacing}{\spaceskip=0pt\relax}
\providecommand{\BIBentryALTinterwordstretchfactor}{4}
\providecommand{\BIBentryALTinterwordspacing}{\spaceskip=\fontdimen2\font plus
\BIBentryALTinterwordstretchfactor\fontdimen3\font minus
  \fontdimen4\font\relax}
\providecommand{\BIBforeignlanguage}[2]{{%
\expandafter\ifx\csname l@#1\endcsname\relax
\typeout{** WARNING: IEEEtran.bst: No hyphenation pattern has been}%
\typeout{** loaded for the language `#1'. Using the pattern for}%
\typeout{** the default language instead.}%
\else
\language=\csname l@#1\endcsname
\fi
#2}}
\providecommand{\BIBdecl}{\relax}
\BIBdecl

\bibitem{zhang2023maxmin}
Y.~Zhang, T.~Zhong, Y.~Wang, J.~Wang, K.~Zheng, and X.~You, ``Max-min fairness
  for uplink {NOMA} systems with finite blocklength,'' \emph{IEEE Transactions
  on Vehicular Technology}, vol.~73, no.~3, pp. 4447--4452, March 2024.

\bibitem{xu2024maxmin}
J.~Xu and B.~Clerckx, ``Max-min fairness and {PHY}-layer design of uplink
  {MIMO} rate-splitting multiple access with finite blocklength,'' \emph{IEEE
  Transactions on Communications}, pp. 1--1, 2024.

\bibitem{liu2019fairness}
Y.~Liu, X.~Chen, L.~X. Cai, Q.~Chen, R.~Gong, and D.~Tang, ``On the fairness
  performance of {NOMA}-based wireless powered communication networks,'' in
  \emph{ICC 2019 - 2019 IEEE International Conference on Communications (ICC)},
  May 2019, pp. 1--6.

\bibitem{wang2020noma}
A.~Wang, L.~Lei, E.~Lagunas, A.~I. Pérez-Neira, S.~Chatzinotas, and
  B.~Ottersten, ``{NOMA}-enabled multi-beam satellite systems: Joint
  optimization to overcome offered-requested data mismatches,'' \emph{IEEE
  Transactions on Vehicular Technology}, vol.~70, no.~1, pp. 900--913, Jan
  2021.

\bibitem{ahsan2020resource}
W.~Ahsan, W.~Yi, Z.~Qin, Y.~Liu, and A.~Nallanathan, ``Resource allocation in
  uplink {NOMA}-{IoT} networks: A reinforcement-learning approach,'' \emph{IEEE
  Transactions on Wireless Communications}, vol.~20, no.~8, pp. 5083--5098, Aug
  2021.

\bibitem{nauman2024efficient}
A.~Nauman, M.~Maashi, H.~K. Alkahtani, F.~N. Al-Wesabi, N.~O. Aljehane,
  M.~Assiri, S.~S. Ibrahim, and W.~U. Khan, ``Efficient resource allocation and
  user association in {NOMA}-enabled vehicular-aided {HetNets} with high
  altitude platforms,'' \emph{Computer Communications}, vol. 216, pp. 374--386,
  2024.

\bibitem{gao2017}
Y.~Gao, B.~Xia, K.~Xiao, Z.~Chen, X.~Li, and S.~Zhang, ``Theoretical analysis
  of the dynamic decode ordering {SIC} receiver for uplink {NOMA} systems,''
  \emph{IEEE Communications Letters}, vol.~21, no.~10, pp. 2246--2249, Oct
  2017.

\bibitem{rinaldi2020}
F.~Rinaldi, H.-L. Maattanen, J.~Torsner, S.~Pizzi, S.~Andreev, A.~Iera,
  Y.~Koucheryavy, and G.~Araniti, ``Non-terrestrial networks in {5G} \& beyond:
  A survey,'' \emph{IEEE Access}, vol.~8, pp. 165\,178--165\,200, 2020.

\bibitem{azari2022}
M.~M. Azari, S.~Solanki, S.~Chatzinotas, O.~Kodheli, H.~Sallouha, A.~Colpaert,
  J.~F. Mendoza~Montoya, S.~Pollin, A.~Haqiqatnejad, A.~Mostaani, E.~Lagunas,
  and B.~Ottersten, ``Evolution of non-terrestrial networks from {5G} to {6G}:
  A survey,'' \emph{IEEE Communications Surveys \& Tutorials}, vol.~24, no.~4,
  pp. 2633--2672, Fourthquarter 2022.

\bibitem{capez2024}
G.~Maiolini~Capez, M.~A. Cáceres, R.~Armellin, C.~P. Bridges, J.~A. Fraire,
  S.~Frey, and R.~Garello, ``On the use of mega constellation services in
  space: Integrating {LEO} platforms into {6G} non-terrestrial networks,''
  \emph{IEEE Journal on Selected Areas in Communications}, vol.~42, no.~12, pp.
  3490--3504, Dec 2024.

\bibitem{ITU-R_S.1528}
{International Telecommunication Union}, ``{Satellite antenna radiation
  patterns for non-geostationary orbit satellite antennas operating in the
  fixed-satellite service below 30 GHz},'' {International Telecommunication
  Union}, Tech. Rep. ITU-R S.1528, 2001.

\bibitem{balanis2016antenna}
C.~A. Balanis, \emph{Antenna Theory: Analysis and Design}, 4th~ed.\hskip 1em
  plus 0.5em minus 0.4em\relax John Wiley \& Sons, 2016.

\bibitem{tse2005fundamentals}
D.~Tse and P.~Viswanath, \emph{Fundamentals of Wireless Communication}, ser.
  Wiley series in telecommunications.\hskip 1em plus 0.5em minus 0.4em\relax
  Cambridge University Press, 2005.

\bibitem{shahab2020}
M.~B. Shahab, R.~Abbas, M.~Shirvanimoghaddam, and S.~J. Johnson, ``Grant-free
  non-orthogonal multiple access for {IoT}: A survey,'' \emph{IEEE
  Communications Surveys \& Tutorials}, vol.~22, no.~3, pp. 1805--1838,
  thirdquarter 2020.

\end{thebibliography}

\end{document}